\documentclass[pra,twocolumn,showpacs,floatfix]{revtex4}

\usepackage[usenames]{color}
\usepackage{graphicx}

\newcommand{\beq}{\begin{equation}}
\newcommand{\eeq}{\end{equation}} 
\newcommand{\beqa}{\begin{eqnarray}}
\newcommand{\eeqa}{\end{eqnarray}}
\newcommand{\ba}{\begin{array}}
\newcommand{\ea}{\end{array}}

\begin{document}

\title{Superfluid Bose-Fermi mixture from weak-coupling to unitarity} 
\author{S. K. Adhikari$^1$\footnote{adhikari@ift.unesp.br; URL 
www.ift.unesp.br/users/adhikari} and 
Luca Salasnich$^2$\footnote{salasnich@pd.infn.it; URL 
www.padova.infm.it/salasnich}} 
\affiliation{$^1$Instituto de F\'isica Te\'orica, UNESP $-$ S\~ao Paulo 
State University, 01405-900 S\~ao Paulo, SP, Brazil\\
$^2$CNR-INFM and CNISM, Unit of Padua,  
Department of Physics ``Galileo Galilei'', 
University of Padua,  35122 Padua, Italy} 

\begin{abstract} 
We investigate the zero-temperature properties of a superfluid 
Bose-Fermi mixture by introducing a set of  
coupled Galilei-invariant nonlinear Schr\"odinger equations 
valid from weak-coupling to unitarity. 
The Bose dynamics is described by a 
Gross-Pitaevskii-type equation including beyond-mean-field corrections
possessing the correct weak-coupling and unitarity limits. The 
dynamics of the two-component Fermi superfluid 
is described by a density-functional  
equation including beyond-mean-field terms 
with correct weak-coupling and unitarity limits. 
The present set of equations is equivalent to the 
equations of generalized superfluid hydrodynamics, 
which take into account also surface effects. The equations 
describe the mixture properly 
as the Bose-Bose repulsive (positive) and Fermi-Fermi attractive (negative)
scattering lengths are varied from zero to infinity in the presence of a
Bose-Fermi interaction.  The present model is tested numerically as the 
Bose-Bose and  Fermi-Fermi  scattering lengths are varied over wide 
ranges covering the weak-coupling to unitarity transition. 
\end{abstract}


\maketitle 

\section{Introduction}

The macroscopic quantum phenomenon of superfluidity has been 
clearly demonstrated in recent experiments with ultracold 
and dilute gases made of alkali-metal atoms \cite{stringa1,stringa2}. 
Both bosonic and fermionic ultracold and dilute 
superfluids can be accurately described by the zero-temperature 
hydrodynamical equations of superfluids 
\cite{stringa1,stringa2,landau1,landau2,leggett}. 

Trapped Bose-Fermi mixtures, with 
Fermi atoms in a single hyperfine state (normal Fermi gas), 
were investigated by various 
authors both theoretically \cite{molmer,nygaard,stoof,pethick,viverit1,
viverit2,das,liu,wang} and experimentally \cite{roati,modugno,ospelkaus}. 
Recently, there was a study of a 
dilute mixture of superfluid bosons and fermions across 
a Feshbach resonance of the Fermi-Fermi scattering length $a_f$, 
obtaining the phase diagram of the mixture in a box \cite{sala-toigo}. 
In the strict one-dimensional case, we found that 
the superfluid Bose-Fermi mixture exhibits phase separation 
and solitons by changing the Bose-Fermi interaction 
\cite{sala-sadhan-flavio,md1}.  

In the present paper we analyze in detail a 3D superfluid Bose-Fermi 
mixture under harmonic trapping confinement when the  
Bose-Bose repulsive (positive) and  Fermi-Fermi attractive 
(negative) 
scattering lengths are varied from very small (weak coupling) to 
infinitely large (strong coupling) values. The variation of these 
scattering lengths from weak to strong coupling is achieved in 
laboratory \cite{fesh} by varying a background magnetic field near a 
Feshbach 
resonance. 

In the first part of the paper we derive Galilei-invariant nonlinear 
Schr\"odinger equations for superfluid Bose and Fermi dynamics,  
valid from weak-coupling to unitarity in each case, which are 
equivalent to the hydrodynamical equations. 
For bosons the equation is a generalized \cite{FP,sala-ggpe,sala-new} 
zero-temperature Gross-Pitaevskii 
(GP) equation \cite{gross} including beyond-mean-field 
corrections to incorporate correctly the effect of bosonic interaction 
for large 
and positive (repulsive) scattering lengths $a_b$. In the ordinary GP 
equation the nonlinear term is linearly proportional to $a_b$ and hence 
highly overestimates the effect of bosonic interaction for large 
positive values of $a_b$. In fact there is a saturation of bosonic 
interaction in the so-called unitarity $a_b\to +\infty$ limit, properly 
taken care of in the present generalized GP equations.  
 For
superfluid fermions the present formulation is based on the 
density-functional (DF) theory \cite{kohn1,kohn2} for fermion pairs
\cite{kim,manini05,adhikari, sala-josephson,sala-new} including
beyond-mean-field corrections to incorporate properly the effect of
fermionic attraction between spin-up and down fermions specially for
large negative (attractive) values of Fermi-Fermi scattering length
$a_f$. The DF theory, in different forms, has already
been applied to the problem of ultracold fermions \cite{kohn,kohnx}.  
The
usual DF equation is valid for the weak-coupling
Bardeen-Cooper-Schreiffer (BCS) limit: $a_f\to -0$. The present 
generalized DF equation correctly accounts for the 
dynamics as the
Fermi-Fermi attraction is varied from the weak-coupling BCS limit to
unitarity: $a_f\to -\infty$. In the unitarity limit there is a
saturation of the effect of fermion interaction, properly taken care of
in the present set of equations, which seems to be appropriate to study
the crossover \cite{bcs-bose} from the weak coupling BCS limit to the
molecular Bose limit.

In the second part of the paper, by including the interaction 
between a boson and fermion pair,  
we obtain a set of coupled equation for superfluid Bose-Fermi dynamics 
in the presence of a Bose-Fermi interaction. The model correctly 
describes the dynamics as the Bose-Bose repulsion and Fermi-Fermi 
attraction are  varied from weak-coupling to unitarity. 
This model is tested numerically 
 as 
the different scattering lengths are varied over wide ranges covering 
the weak-coupling to unitarity transition for a superfluid Bose-Fermi 
mixture. 

In Sec. II we present the hydrodynamical equations for bosons and 
fermions and present the appropriate bulk chemical potentials valid in 
the weak-coupling limit as well as possessing the saturation in the 
strong-coupling unitarity limit consistent with the constraints of 
quantum mechanics. In Sec. III we derive mean-field equations for bosons 
and fermions consistent with the hydrodynamical  equations for a large 
number of atoms. Next we derive the present  model for interacting 
Bose-Fermi superfluid introducing a contact 
interaction between bosons and fermions. 
In Sec. IV we perform numerical calculations for densities and chemical 
potentials of  a trapped  Bose-Fermi superfluid   
mixture to show the advantage of the present model.
Finally, in Sec. V we present  a brief summary and conclusion.  

\section{Superfluid hydrodynamics for bosons and fermions} 

At zero temperature, for a large number of atoms, statical and dynamical 
collective 
properties of bosonic and fermionic superfluids 
are expected to be properly described by the 
hydrodynamical equations of superfluids \cite{stringa1,stringa2}. 
For bosons the hydrodynamical equations
are given by \cite{stringa1,landau1} 
\beqa 
{\partial \over \partial t} n_b &+& 
\nabla \cdot \left( n_b {\bf v}_b \right) = 0 \; , 
\label{hy-1}
\\
m_b {\partial \over \partial t} {\bf v}_b &+& 
\nabla \left[ {1\over 2} m_b v_b^2 + U_b + \mu_b(n_b,a_b) 
\right] = 0 \; ,  \label{hy-2}
\eeqa 
where $U_b({\bf r})$ is the external potential acting on bosons, 
$m_b$ is the mass of a bosonic atom, 
$n_b({\bf r},t)$ is the local density of bosons,  
and ${\bf v}_b({\bf r},t)$ is the local superfluid velocity 
\cite{stringa1,stringa2,landau1}. 
The total number of bosons is given by 
\beq 
N_b = \int n_b({\bf r},t) \ d^3{\bf r} \; ,  
\eeq 
and the nonlinear term $\mu_b(n_b,a_b)$ is the bulk 
chemical potential of the bosonic system 
with $a_b$ the Bose-Bose  scattering length. 
Equation (\ref{hy-1}) is the equation of continuity of hydrodynamic flow,  
while Eq. (\ref{hy-2}) is the equation of 
conservation of the linear momentum. Equation  (\ref{hy-2}) establishes 
the irrotational nature of the superfluid motion: $\nabla \wedge {\bf 
v}_b = 0$, 
meaning that the velocity ${\bf v}_b$ can be written as the 
gradient of a scalar field. Equations  (\ref{hy-1}) and (\ref{hy-2}) 
differ from the corresponding equations holding in the 
collisional regime of a non-superfluid system because of 
the irrotationality constraint. 

For superfluid fermions, the fundamental entity governing the 
superfluid flow is the Cooper pair. In the case of a two-component
(spin up and down) 
superfluid Fermi system one can introduce 
the local density of pairs $n_p({\bf r},t)=n_f({\bf r},t)/2$, 
where $n_f({\bf r},t)$ is the total local density of fermions. 
The hydrodynamical equations of a fermionic superfluid 
\cite{stringa2,landau1,landau2,leggett} can be written as 
\beqa 
{\partial \over \partial t} n_p &+& 
\nabla \cdot \left( n_p {\bf v}_p \right) = 0 \; , 
\label{hyp-1}
\\
m_p {\partial \over \partial t} {\bf v}_p &+& 
\nabla \left[ {1\over 2} m_p v_p^2 + U_p + \mu_p(n_p,a_f) 
\right] = 0 \; ,  
\label{hyp-2}
\eeqa 
where the nonlinear term $\mu_p(n_p,a_f)$ is the bulk 
chemical potential of pairs with $a_f$ the attractive 
Fermi-Fermi scattering 
length. Here
$m_p$ is the mass of a pair, that is twice the mass $m_f$ 
of a single fermion, i.e. $m_p=2m_f$. 
In addition, 
the trap potential $U_p({\bf r})$ is twice the 
trap potential $U_f({\bf r})$ acting on a single fermion, 
i.e. $U_p({\bf r})=2U_f({\bf r})$. 
Note that the chemical potential $\mu_p(n_p,a_f)$ 
is twice the total chemical potential $\mu_f(n_f,a_f)$ of the 
Fermi system, i.e. $\mu_p(n_p,a_f)=2\mu_f(2n_p,a_f)$. 
The total number $N_f$  of fermions is 
\beq 
N_f = 2 N_p = 2 \int n_p({\bf r}) \ d^3{\bf r}= \int n_f({\bf r}) \ 
d^3{\bf r} \; ,  
\eeq 
where $N_p$ is the number of pairs. 

It is important to stress that Eqs. (\ref{hy-1}) and (\ref{hy-2}) 
are formally similar to Eqs. (\ref{hyp-1}) and (\ref{hyp-2}),  
but the quantities involved are strongly different. 
In particular, the bulk chemical potential of 
bosons is completely different from the bulk chemical 
potential of Fermi pairs \cite{stringa1,stringa2,sala-ideal}. 

In the full crossover from the small-gas-parameter regime to the 
large-gas-parameter regime we use the following expression for the bulk  
chemical potential of the Bose superfluid \cite{sala-ggpe}
\beq \label{cpbose}
\mu_b(n_b,a_b) = \frac{\hbar^2}{m_b}n_b^{2/3}f(n_b^{1/3}a_b) \; , 
\eeq
where 
\beq \label{fx}
f(x) =  4\pi  \frac{x+\alpha x^{5/2}}{1+\gamma x^{3/2}+
\beta x^{5/2}} \; .
\eeq
In the small-gas-parameter ($x\to 0$) regime, Eq.
(\ref{fx}) becomes 
\beq\label{fx2} f(x)= 4\pi \left[
x+(\alpha-\gamma)x^{5/2}+...\right], \eeq 
which is the analytical result
for bulk chemical potential found by Lee, Yang, and 
Huang \cite{LY,LYH}
in this limit, provided that $(\alpha-\gamma)=32/(3\sqrt \pi)$.
In Eq. (\ref{fx}) we shall also choose $\beta=4 \pi \alpha /\eta$, with 
$\eta=22.22$; this will make the bulk chemical potential  (\ref{cpbose})
satisfy the correct unitarity limit $\mu_b(n_b,a_b) = 22.22 \hbar^2 
n_b^{2/3}/m_b$ as established by Cowell { \it et al.} \cite{cowell}.
Hence Eqs. (\ref{cpbose}) with (\ref{fx}) can be made to satisfy the 
correct weak-coupling and unitarity limits. Next we need to determine 
the 
constants $\alpha$ and $\gamma$ consistent with 
$(\alpha-\gamma)=32/(3\sqrt \pi)$. We consider the following three 
choices for 
$\alpha$ and $\gamma$ from a possible many choices:
$\alpha=32\nu /(3\sqrt \pi)$,
$\gamma=32(\nu-1)/(3\sqrt \pi)$
with (a) $\nu=1.05$, (b) $\nu=1.1$, and (c) $\nu=1.15$.
All three choices are consistent with the unitarity and the 
Lee-Yang-Huang limits \cite{LY,LYH}.  We shall present a critical 
numerical study of the three choices in the next section.

In the full crossover from BCS regime ($a_f \to -0$) to 
unitarity ($a_f\to -\infty$) we 
use the following 
expression for the bulk chemical potential of the Fermi superfluid 
\cite{adhik} 
\beq \label{cpfermi}
\mu_p(n_p,a_f) = 2\frac{\hbar^2}{m_p}(6\pi^2 
n_p)^{2/3}g(2^{1/3}n_p^{1/3}a_f)
\eeq
where 
\beq \label{gx}
g(x) =  1+  \frac{\delta x}{1-\kappa x} \; ,
\eeq 
where $\delta$ and $\kappa$ are fitting parameters. 
The close 
agreement of the interpolation function $g(x)$ for
negative $x$ values with the results of Monte Carlo calculation of
energy of a  uniform gas of Fermi superfluid   with the parameters  
$\delta=4\pi/(3 \pi^2)^{2/3}$, $\kappa=
\delta /(1-\zeta)$, and  $\zeta=0.44$ \cite{th1,th2}
was established in Ref. 
\cite{adhik}. 
The inclusion of the lowest-order term $g(x)=1$ in Eq. (\ref{cpfermi})
leads to the bulk chemical potential in the absence of Fermi-Fermi 
interaction ($a_f=0$) of a uniform gas. The next-order 
term $g(x)=1+\delta x$
include known analytical result in the small-gas-parameter regime as 
obtained by Lee and Yang \cite{LY,heisel} {and 
by Galitskii 
\cite{galit}.  }
The model (\ref{cpfermi}) with (\ref{gx})  provides a smooth 
interpolation between the bulk chemical potential in the BCS limit  
$\mu_p(n_p,a_f)=2\hbar^2 (6\pi^2 
n_p)^{2/3}/m_p+{16}\pi n_p \hbar^2 a_f/m_p$ \cite{LY}
for  small $n_p^{1/3}a_f$ values and that in the unitarity limit 
\cite{adhik}
$\mu_p(n_p,a_f)= 2\hbar^2 (6\pi^2
n_p)^{2/3}\zeta /m_p$ for large $n_p^{1/3}a_f$ values.  
Manini and Salasnich \cite{manini05} and Kim and Zubarev 
\cite{kim}
also  proposed 
similar interpolation formulas on the basis of Monte Carlo data of a 
uniform Fermi superfluid. 

Here we deal with a Fermi gas in a spherical harmonic trap rather than 
a uniform Fermi gas. By a direct comparison with the results of Monte 
Carlo calculation of  a superfluid Fermi gas in a spherical harmonic 
trap, we shall show  in Sec. III that Eqs. (\ref{cpfermi}) and 
(\ref{gx}) still 
present a good approximation to the bulk chemical potential of the 
system, but now with a slightly different    value of the parameters. 

\begin{figure}[tbp]
\begin{center}
{\includegraphics[width=.8\linewidth,clip]{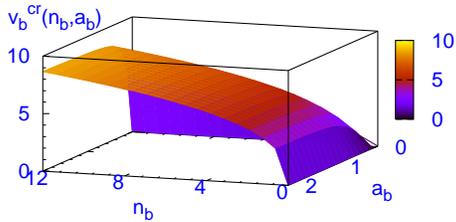}}
\end{center}
\caption{(Color online) The critical sound velocity $v_b^{cr}$
of a uniform 
Bose superfluid 
as a function of Bose-Bose scattering length $a_b$ and 
density $n_b$.  }
\label{fig1}
\end{figure}

The hydrodynamical equations are valid to describe 
equilibrium properties and dynamical properties of 
long-wavelength for both bosons and fermions. 
In particular, one can introduce \cite{stringa1,stringa2} a healing (or 
coherence) 
length such that the transport phenomena under investigation 
must be characterized by a length scale much larger than 
the healing length. 
As suggested by Combescot, Kagan and Stringari \cite{combescot}, 
the healing length can be defined for bosons as 
\beq 
\xi_b = {\hbar\over m_b v_b^{cr}} \; , 
\label{healing}
\eeq
where $v_b^{cr}$ is the Landau critical velocity above 
which the system gives rise to energy dissipation. 
This critical velocity coincides 
with the first sound velocity and is given by  
\beq 
v_b^{cr} = \sqrt{{n_b\over m_b}{\partial \mu_b \over \partial n_b}} \; . 
\label{v-cr-b}
\eeq
In Fig. \ref{fig1}  we plot the critical sound velocity $v_b^{cr}$ 
of a uniform Bose superfluid for 
the present $\mu_b$ given by Eqs. (\ref{cpbose}) and (\ref{fx}) with 
the parameter $\nu =1.1$. 
The 
interesting feature of this plot is that  $v_b^{cr}=0$, when either 
$n_b$ or $a_b$ is zero. However, $v_b^{cr}$ 
increases monotonically  with $n_b$, whereas $v_b^{cr}$ attains a 
constant value as $a_b$  increases. These features are present in the 
plot of Fig. \ref{fig1}. However, the critical sound velocity 
$v_b^{cr}$ calculated from $\mu_b$ given by Eqs. (\ref{cpbose}) 
and (\ref{fx2})  increases monotonically  with both $n_b$ 
and $a_b$.

For superfluid fermions the healing length of Cooper pairs 
can be defined as 
\beq 
\xi_p = {\hbar \over m_p v_p^{cr}} \; , 
\eeq
where the critical velocity $v_p^{cr}$
is related to the breaking of pairs through the formula 
\beq 
v_p^{cr} = \sqrt{ \sqrt{\mu_p^2 + |\Delta|^2}-\mu_p \over m_p} \; , 
\label{v-cr-f}
\eeq
where $|\Delta|$ is the energy gap \cite{stringa2,combescot}. 
We notice that in the deep BCS regime of weakly interacting 
attractive Fermi atoms 
(corresponding to $|\Delta| \ll \mu_p$) Eq. (\ref{v-cr-f}) 
approaches the exponentially small value 
{$v_p^{cr}=|\Delta|/\sqrt{m_p \mu_p/2}$.} 

\section{Nonlinear Schr\"odinger equations} 

The bosonic superfluid can be described by a GP \cite{gross,stringa1} 
complex order parameter $\Psi_b({\bf r},t)$ given by 
\beq 
\label{eq1}
\label{psi-b}
\Psi_b({\bf r},t) =\sqrt{n_b({\bf r},t)} \ e^{i\theta_b({\bf r},t)} \ . 
\eeq 
The probability current density $\vec j({\bf r},t)$ is given by
\beq\label{eq2}
\vec j\equiv n_b{\bf v_b}= \frac{\hbar}{2im_b}(\Psi_b ^*\nabla
\Psi_b-\Psi _b\nabla \Psi_b^*).
\eeq
Equations (\ref{eq1}) and (\ref{eq2}) relate the phase  $\theta_b({\bf 
r},t)$ 
to the superfluid velocity field  ${\bf v}_b$
by the formula \cite{stringa1}: 
\beq  
{\bf v}_b = {\hbar \over m_b} \nabla \theta_b \; . 
\label{v-general-b} 
\eeq 
The bosonic order parameter $\Psi_b({\bf r},t)$ is, 
apart from a  normalization \cite{landau2,leggett}, 
nothing but the condensate wave function 
\beq 
\Xi_b({\bf r},t) = 
\langle{\hat \psi}({\bf r},t)\rangle \; , 
\eeq
that is the expectation value of the bosonic field operator 
${\hat \psi}({\bf r},t)$ \cite{landau2,sala-new}. 
The order parameter $\Psi_b({\bf r},t)$ 
satisfies the following generalized GP 
nonlinear Schr\"odinger equation \cite{stringa1} 
\beq
i \hbar {\partial \over \partial t} \Psi_b 
= \left[ - {\hbar^2 \over 2m_b} \nabla^2 + 
U_b + \mu_b(n_b,a_b) \right] \Psi_b \; ,  
\label{nlse-b} 
\eeq
where the bulk chemical potential is given by Eq. (\ref{cpbose}).
The use of $f(x)=4\pi x$ in Eq. (\ref{cpbose}) 
leads to the GP
equation and the use of the two leading terms of Eq. (\ref{fx2}), e.g., 
\beq\label{MGP}
f(x)=4\pi \left[x+{32\over 3\sqrt \pi}x^{5/2}\right]
\eeq 
in
Eq. (\ref{cpbose}) leads to the modified GP (MGP) equation   
suggested by
Fabrocini and Polls \cite{FP}, which includes correction to the GP 
equation for medium values of Bose-Bose scattering length $a_b$.
The use of Eq. (\ref{fx}) in Eq. (\ref{cpbose}) leads to a nonlinear 
Schr\"odinger equation valid from the weak-coupling to the  unitarity 
limit and is called the unitary Schr\"odinger (US) equation for bosons
\cite{sala-ggpe}.

\begin{figure}[tbp]
\begin{center}
{\includegraphics[width=.8\linewidth,clip]{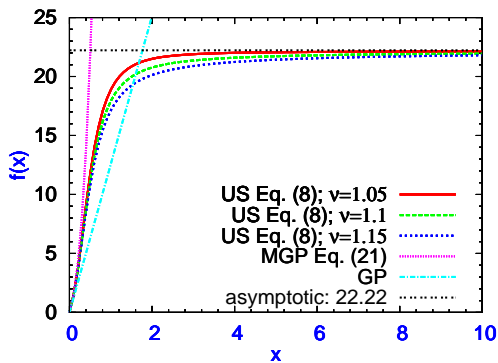}}
\end{center}
\caption{(Color online) The Bose interpolation function $f(x)$ given by  
Eq. (\ref{fx}) with $\alpha=32\nu/(3\sqrt \pi), \gamma 
=32(\nu-1)/(3\sqrt\pi),$ and $ \beta =4\pi \alpha/22.22$ with $\nu 
=$ (a) 1.05, (b) 1.1, and  (c) 1.15. The results of MGP [Eq. 
(\ref{MGP})] and GP
[$f(x)=4\pi x$] 
models 
together 
with the asymptotic value are also given. 
}
\label{fig2}
\end{figure}

In Fig. \ref{fig2} we plot the various possibilities for the
interpolation function $f(x)$; e.g., those corresponding to the US
equation (\ref{fx}) with (a) $\nu=1.05$, (b)  $\nu=1.1$, and 
(c) $\nu=1.15$ [as suggested after Eq. (\ref{fx2})],  the 
MGP
equation (\ref{MGP}), the GP equation $f(x)=4\pi x$. The asymptotic
limit $\lim_{x\to \infty}f(x)=22.22$ is also shown.  
All three choices (a), (b), and (c)  
represent smooth  interpolation  
of $f(x)$ between  small and large $x$ values. For small 
$x$ this
choice agrees with the MGP result (\ref{MGP}) and for large $x$ with the
asymptotic result.  To test these three choices we actually solve 
the US Eq. (\ref{nlse-b}), by imaginary time propagation after a 
Crank-Nicholson 
discretization \cite{num,sala-numerics,cn}
(detailed in the beginning of Sec. IV),
for the trap parameters of a possible 
experimental set up for $^{87}$Rb atoms in a spherical trap and compare 
the results for energy 
with those obtained by Blume and Greene \cite{bg} by diffusion Monte 
Carlo (DMC) 
method. Actually, we solved Eqs. (20) and (7) with $n_b$ normalized to 
$(N_b-1)$ in place of Eq. (3), appropriate for a small number of bosons, 
cf. Eq. (2) of Ref. [48].  
The energy of the system is calculated through a numerically 
constructed energy functional from the present bulk chemical potential. 
The calculation is performed for a boson-boson scattering length $a_b= 
0.433l$ where $l$ is the harmonic oscillator length for various $N_b$ 
and the results for energy are tabulated in Table \ref{table1}. 
From this table we find that the present US models always provide  
a much better approximation to the energies than the GP and MGP models. 
More results of energy for larger number of bosons may help in fixing 
the parameters of the present model more accurately. In the present 
paper  we shall use the US model (b) with $\nu =1.1$.

\begin{table}[!ht]
\begin{center}
\caption{Ground state energies  for different 
number $N_b$ of bosonic 
atoms in a spherical trap and for $a_b=0.433$  from 
a solution of GP 
equation, MGP equation 
and US equation for $\nu$ = (a) 1.05, (b) 1.1 and (c) 1.15. The results 
for energies 
obtained with two potentials of   DMC 
calculation \cite{bg} are quoted for comparison.
Length and energies are 
expressed in 
oscillator units.   
}
\label{table1}
\begin{tabular}{|r|r|r|r|r|r|r|}
\hline
$N_b$ &   DMC &
GP & MGP & US (a)&US(b)&US(c) \\
\hline
3 & 5.553(3); 5.552(2)  &  5.329 &   5.611 & 5.570  &5.564 & 5.558\\
5 & 10.577(2);  10.574(4)  &  9.901&    10.772  & 10.629 &10.608 &10.588  \\
10 & 26.22(8); 26.20(8)  &   23.61  &  26.84 &     26.24 &26.16  &26.07 \\
 20 & 66.9(4); 66.9(1)  &   57.9   &   68.5&     66.38 &66.07 & 65.77\\
\hline
\end{tabular}
\end{center}
\end{table}

The relationship between Eq. (\ref{nlse-b}) and 
the hydrodynamical  equations (\ref{hy-1}) and (\ref{hy-2}) can be 
established by inserting Eq. (\ref{psi-b}) into Eq. (\ref{nlse-b}). 
After some straightforward algebra equating the real and imaginary parts  
of both sides and taking into account Eq. (\ref{v-general-b}), 
one finds two generalized hydrodynamical equations \cite{stringa1}
\beqa 
\label{eq3}
{\partial \over \partial t} n_b 
&+&    \nabla \cdot \left( n_b {\bf v}_b \right) = 0 \; , 
\\ 
m_b {\partial \over \partial t} {\bf v}_b &+&
\nabla \biggr[ - 
{\hbar^2\over 2m_b} 
{\nabla^2 \sqrt{n_b}\over \sqrt{n_b}} + 
{ m_b v_b^2 \over 2}+  \ U_b \nonumber \\ &+& 
 \ \mu_b(n_b,a_b) \biggr] = 0 \; , \label{eq4}  
\eeqa 
which include the quantum pressure term 
\beq
T_b^{QP}=-
{\hbar^2\over 2m_b}
{\nabla^2 \sqrt{n_b}\over \sqrt{n_b}} \; , 
\label{qp} 
\eeq 
which depends explicitly on the reduced 
Planck constant $\hbar$. Neglecting the quantum pressure term, 
one gets from Eqs. (\ref{eq3}) and (\ref{eq4}) the classical 
hydrodynamical equations 
(\ref{hy-1}) and (\ref{hy-2}). Equation (\ref{eq3}) is the continuity 
equation whereas Eq. (\ref{eq4})  establishes the irrotational nature of 
hydrodynamic flow. 

\begin{figure}[tbp]
{\includegraphics[width=.8\linewidth,clip]{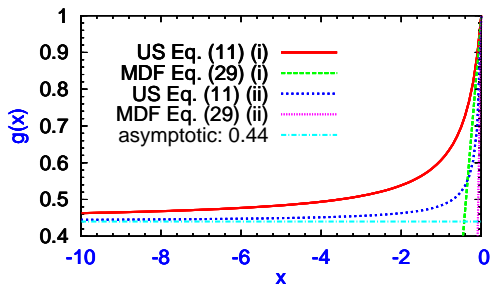}}
\caption{(Color online) The Fermi interpolation function $g(x)$
given by Eq. (\ref{gx}) with (i)  $\delta = $ $4\pi/(3\pi^2)^{2/3},$ 
and (ii)  $20\pi/(3\pi^2)^{2/3}$ and $\kappa=\delta/0.56$. The MDF 
results given by Eq. (\ref{MDF}) and the asymptotic value are also 
shown.}
\label{fig3}
\end{figure}

Similarly, the fermionic superfluid can be described by a 
DF \cite{kohn} complex order parameter $\Psi_p({\bf r},t)$ of 
pairs, 
with a modulus and a phase $\theta_p({\bf r},t)$ such that  
\beq 
\label{psi-p}
\Psi_p({\bf r},t) =\sqrt{n_p({\bf r},t)} \ e^{i\theta_p({\bf r},t)} 
\;  . 
\eeq
The expression for the probability current density of fermions leads to 
the following expression for 
the velocity field ${\bf v}_p$
of pairs \cite{stringa2}:  
\beq  
{\bf v}_p = {\hbar \over m_p} \nabla \theta_p \; . 
\label{v-general-p}
\eeq 
Note that the mass of a pair $m_p$, and not that of a fermion, 
appears in the denominator of the phase-velocity relation.
The fundamental entity responsible for superfluid flow has the mass 
$m_p$ of 
a pair of fermions. 
For paired fermions in the superfluid state, 
the  order parameter is, apart from a  
normalization \cite{landau2,leggett,sala-odlro}, 
the condensate wave function of the center of mass 
of the Cooper pairs:  
\beq 
\Xi_p({\bf r},t) = 
\langle {\hat \psi}_{\uparrow}({\bf r},t) 
{\hat \psi}_{\downarrow}({\bf r},t) \rangle \; , 
\eeq 
that is the average of pair operators, 
with ${\hat \psi}_{\sigma}({\bf r},t)$ the fermionic 
field operator with spin component $\sigma=\uparrow,\downarrow$ 
\cite{landau2,sala-new}. 
The order parameter $\Psi_p({\bf r},t)$
satisfies the following zero-temperature
nonlinear DF Schr\"odinger equation
\cite{kim,manini05,sala-josephson,sala-new,adhik}
\beq
i \hbar {\partial \over \partial t} \Psi_p 
= \left[ - {\hbar^2 \over 2m_p} \nabla^2 + 
 \ U_p +  \ \mu_p(n_p,a_f) \right] \Psi_p \; ,  
\label{nlse-p}
\eeq 
where the bulk chemical potential $\mu_p(n_p,a_f)$
 is given by Eq. (\ref{cpfermi}). The use of $g(x)=1$ in $ 
\mu_p(n_p,a_f)$ of Eq. (\ref{cpfermi})
 leads to the 
lowest order DF equation. If we use the next order solution  
of Eq. (\ref{gx}) in Eq. (\ref{cpfermi}), e. g.,
\beq
g(x)=1+\delta x, 
\label{MDF} 
\eeq
we obtain a modified density-functional (MDF) equation with corrections 
for medium values Fermi-Fermi  scattering length $a_f$. The use of Eq. 
(\ref{gx}) in Eq. (\ref{cpfermi}) leads to a DF  
equation valid in the weak-coupling BCS to unitarity limit and the 
resultant nonlinear equation (\ref{nlse-p}) will be called the unitary 
Schr\"odinger 
(US) equation for fermion pairs.

\begin{figure}[tbp]
{\includegraphics[width=.8\linewidth,clip]{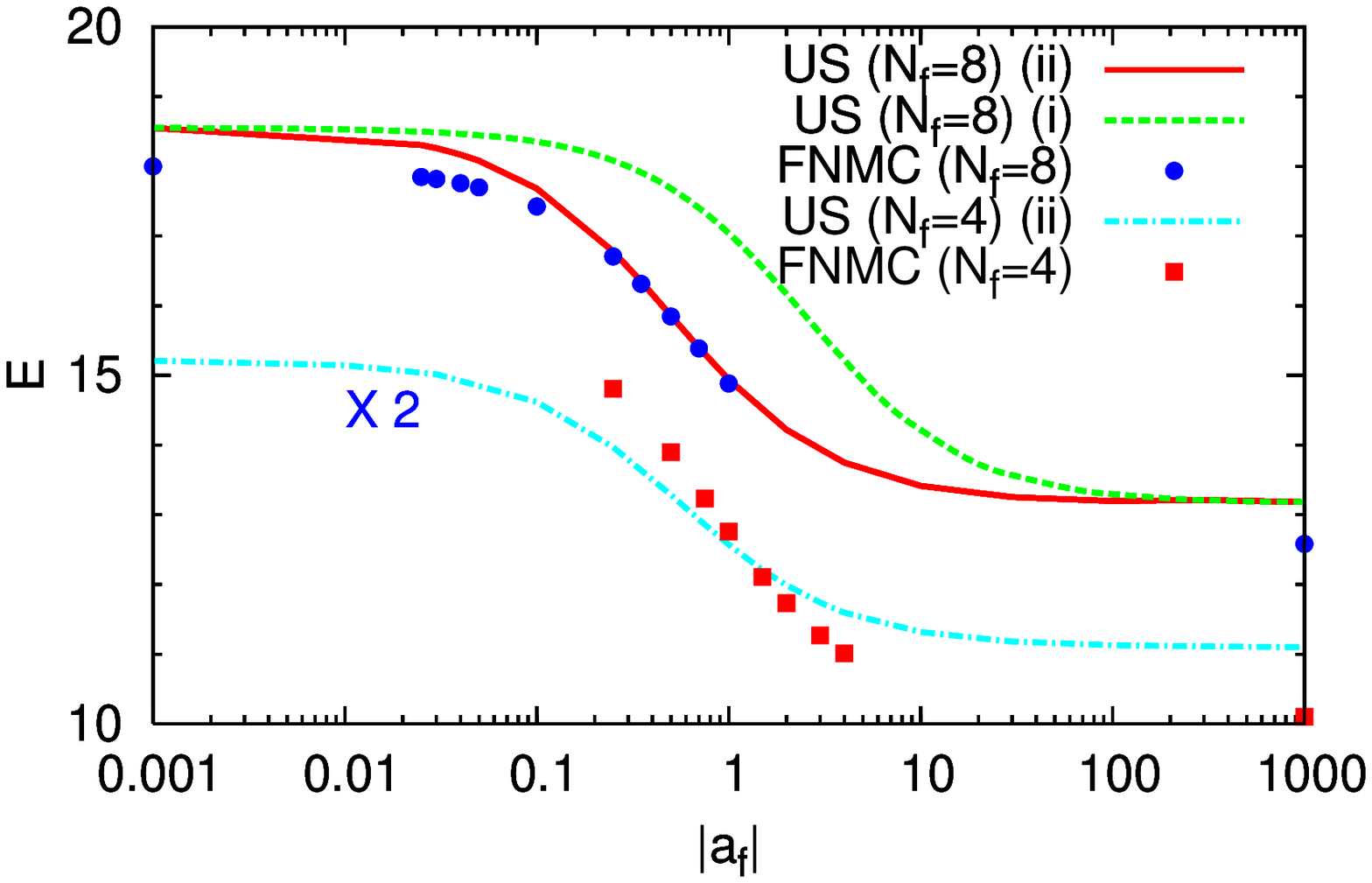}}
{\includegraphics[width=.8\linewidth,clip]{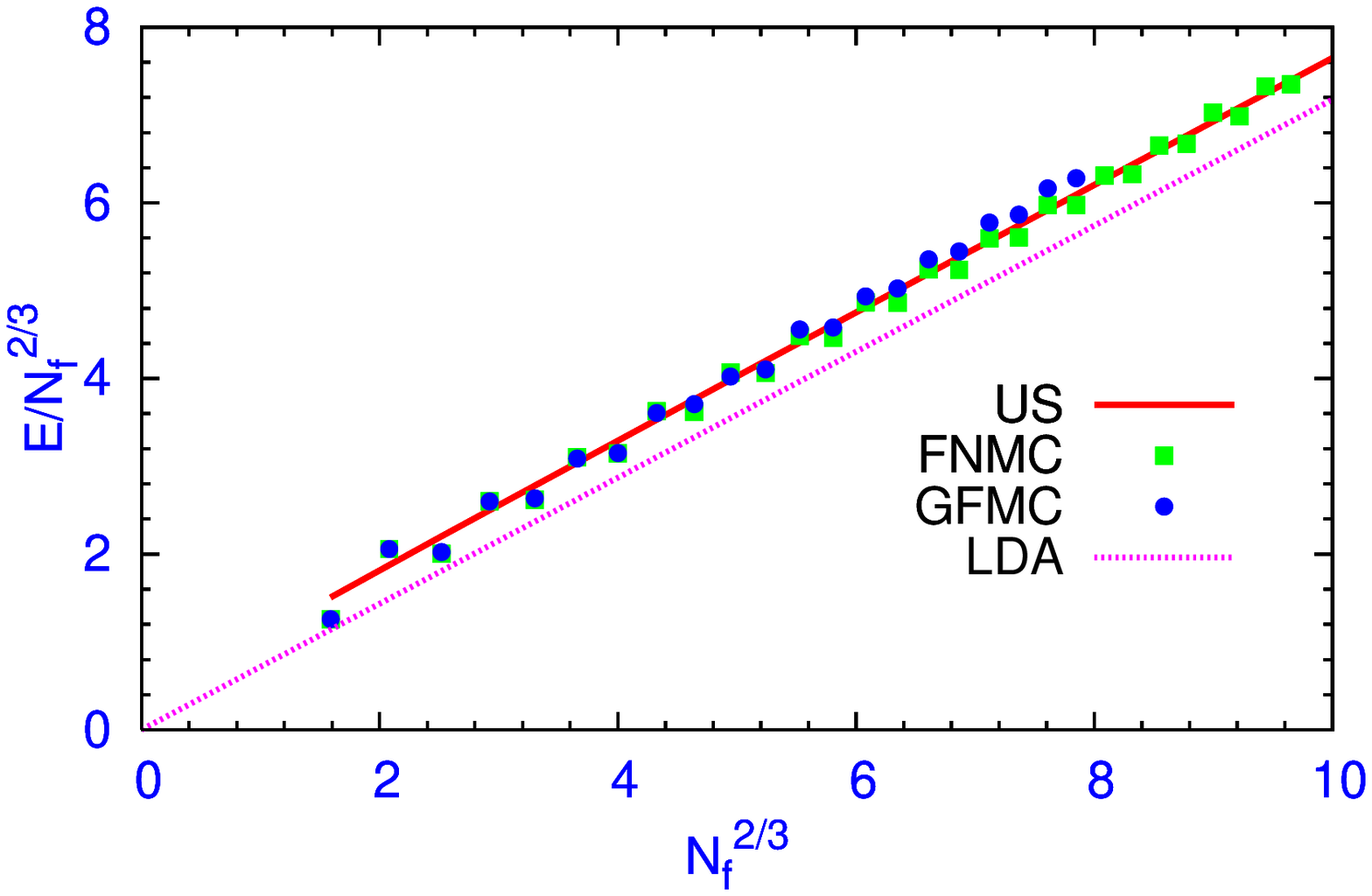}}
\caption{(Color online) (a) Energy of a superfluid Fermi gas in a 
spherical trap in oscillator units vs. Fermi-Fermi scattering length 
$a_f$ also in oscillator units obtained from the solution of Eq. 
(\ref{nlse-p}) for the choices  $\delta=$  (i) $ 4\pi/(3\pi^2)^{2/3}$ 
and 
(ii) $ 20\pi/(3\pi^2)^{2/3}$ and from the  fixed node Monte Carlo
(FNMC) calculation \cite{FNMC,FNMC2}. (b) Energy 
of 
a superfluid Fermi gas in a
spherical trap in oscillator units vs.  number of atoms $N_f$ in the 
unitarity limit $a_f \to -\infty$ obtained from a solution of Eq. 
(\ref{nlse-p}), from FNMC \cite{FN}
and  
Green function Monte Carlo  (GFMC)
calculations \cite{GFMC}, and from local density approximation (LDA). 
}  
\label{fig4}
\end{figure}

In Fig. \ref{fig3} we plot the various possibilities for the
interpolation function $g(x)$; e. g., that corresponding to Eq.  
(\ref{gx}), the MDF
equation (\ref{MDF}) for two different values of $\delta$:  
 (i)  $4\pi/(3\pi^2)^{2/3}$ and (ii) $20\pi/(3\pi^2)^{2/3}$. 
The parameter 
$\kappa$ is always taken as $\kappa=\delta/(1-\zeta),$ $ \zeta=0.44$
\cite{th1,th2}. 
The 
asymptotic
limit $\lim_{x\to \infty}g(x)=0.44$ is also shown in Fig. \ref{fig3}.  
The expression 
(\ref{gx}) represents a good
approximation of $g(x)$ for small and large $|x|$. For small $|x|$ this
choice agrees with the MDF result (\ref{MDF}) and for large $|x|$ with 
the asymptotic result for both choices of $\delta$. 

To test the above two choices (i) and (ii) of the parameter  $\delta$ in 
Eq. 
(\ref{gx}) for a Fermi superfluid in a spherical trap, we calculate the 
energy of the system for different values of the scattering length $a_f$ 
and number of atoms $N_f$ by solving directly Eq.
(\ref{nlse-p}) 
by imaginary time propagation after a
Crank-Nicholson
discretization \cite{num,sala-numerics,cn}
(detailed in the beginning of Sec. IV).
We also  compare the results with those obtained 
by the  fixed-node  Monte Carlo calculation (FNMC) 
\cite{FN,FN2,FNMC,FNMC2}. 
The results of our investigation are shown in Figs. 
\ref{fig4}.   In Fig. \ref{fig4} (a) we plot energy 
$E$ vs. $|a_f|
$ for 
$N_f=4$ and 8 obtained from a solution the US equation (\ref{nlse-p}) 
with the choices (i)  $4\pi/(3\pi^2)^{2/3}$   and (ii)  
$20\pi/(3\pi^2)^{2/3}$  for $\delta$ and the FNMC data 
of Ref. \cite{FNMC,FNMC2}.  The present hydrodynamical formulation is 
expected to be good for for a large number of fermions [see, Fig.  
\ref{fig4} (b)], yet for a small number $N_f=4$ the result is 
reasonable.  
In Fig. 
\ref{fig4} (b)  we 
plot energy $E/N_f^{2/3}$ vs. $N_f^{2/3}$ at 
unitarity $a_f \to -\infty$ obtained from a solution of the US equation 
(\ref{nlse-p}) with choice (ii) $20\pi/(3\pi^2)^{2/3}$  for $\delta$.
In this limit both choices of $\delta$ lead to the same energy. 
The results of FNMC  \cite{FN} and Green function Monte 
Carlo (GFMC) \cite{GFMC} 
calculations as well as of LDA 
are also shown in Fig. \ref{fig4} (b). The LDA result is analytically 
known in this case as $E(N)=(3N_f)^{4/3}\sqrt \xi/4, \xi = 0.44$  
\cite{kohnx}. 
We plot $E/N_f^{2/3}$ vs. $N_f^{2/3}$ in Fig. \ref{fig4} (b) 
because of the 
linear correlation between these two variables explicit in the 
LDA result.
From Figs. \ref{fig4} we find 
that 
the results with choice (ii)
 $20\pi/(3\pi^2)^{2/3}$ 
of $\delta $  agree well with the Monte Carlo data in both cases and 
this choice will be used in the present study.  
The Monte Carlo data clearly favors the present model over the LDA.

Again, inserting Eq. (\ref{psi-p}) into Eq. (\ref{nlse-p}),
after some straightforward algebra equating the real and imaginary parts
and taking into account Eq. (\ref{v-general-p}), 
we find two generalized hydrodynamical equations for the Fermi superfluid 
\beqa 
\label{eq3a}
{\partial \over \partial t} n_p 
&+&    \nabla \cdot \left( n_p {\bf v}_p \right) = 0 \; , 
\\ 
m_p {\partial \over \partial t} {\bf v}_p &+&
\nabla \biggr[ - 
{\hbar^2\over 2m_p} 
{\nabla^2 \sqrt{n_p}\over \sqrt{n_p}} + 
{ m_p v_p^2 \over 2}+  \ U_p \nonumber \\ &+& 
 \ \mu_p(n_p,a_f) \biggr] = 0 \; , 
\label{eq4a}  
\eeqa 
which include the quantum pressure term 
\beq
T_p^{QP}=-{\hbar^2\over 2m_p}
{\nabla^2 \sqrt{n_p}\over \sqrt{n_p}} \; , 
\label{qpa} 
\eeq 
involving $\hbar$. Neglecting this term leads to the classical 
hydrodynamical 
equations (\ref{hyp-1}) and (\ref{hyp-2}) of superfluid Fermi pairs. 
This specific quantum pressure term is a consequence of the proper
phase-velocity relation (\ref{v-general-p}) 
with the factor $m_p$ in the denominator  \cite{stringa2}. 
Had we taken a different mass factor in the denominator, 
e.g. $m_f$, a different quantum pressure term would have emerged. 
The present choice is physically motivated by Galilei invariance 
\cite{stringa2,sala-josephson,sala-new} and leads to energies of 
trapped Fermi superfluid in close agreement \cite{adsa} 
with those obtained  \cite{FN,GFMC} from the Monte Carlo methods. 

We stress that, from the point of view of DF theory, 
the quantum pressure terms for superfluid 
bosons and fermions correspond to gradient correction 
(also called surface corrections) 
to the local density approximation (LDA), where LDA gives exactly 
the classical hydrodynamical equations obtained by setting gradient 
kinetic-energy
term to zero. 
The gradient term, which can also be seen as 
a next-to-leading contribution in a momentum expansion 
to the  classical superfluid hydrodynamics \cite{son}, 
takes into account corrections to the 
kinetic energy due to spatial variations in 
the density of the system, and it is essential to have regular 
behavior at the surface of the cloud where the density 
goes to zero \cite{tosi}. In the case of superfluid 
bosons, the quantum-pressure term, Eq. (\ref{qp}), is exactly 
the one of the familiar cubic GP equation 
\cite{stringa1,gross}. 
In the case of superfluid fermions the gradient 
term is consistent with the motion of paired fermions of mass 
$m_p=2m_f$. For normal fermions various 
authors have proposed and used different gradient terms 
\cite{von,tosi,zaremba,sala-gradient,adhikari2}. For superfluid fermions 
we are
using the familiar von Weizs\"acker term \cite{von} in the case of 
pairs. In fact, in our approach, the explicit 
form of the quantum-pressure term, Eq. (\ref{qpa}),  
is strictly determined by the Galilei-invariant relation, 
Eq. (\ref{v-general-p}), between phase and superfluid velocity 
\cite{stringa2,sala-josephson,sala-new}. 
Finally, we observe that Eq. (\ref{qpa}) 
is practically the same as that recently obtained in Ref. \cite{save-us} 
in the case of a superfluid Fermi gas at unitarity 
using an $\epsilon$ expansion around $d=4-\epsilon$ spatial dimensions. 

The generalized hydrodynamical equations  (\ref{eq3}) 
and (\ref{eq4}) for bosons 
and (\ref{eq3a}) and (\ref{eq4a}) for fermions are thus completely 
equivalent to the respective nonlinear Schr\"odinger equations (\ref{nlse-b}) 
and (\ref{nlse-p}). 
In the limit of zero velocity fields, 
${\bf v}_b, {\bf v}_p \to 0$,  Eqs. (\ref{eq4}) and 
(\ref{eq4a})  yield  the stationary versions of Eqs. (\ref{nlse-b}) and 
(\ref{nlse-p}) describing the superfluid densities: 
\beqa\label{eq10}
 \mu_{b0} \sqrt {n_b}
= \left[ - {\hbar^2 \over 2m_b} \nabla^2 +
 \ U_b +  \ \mu_b(n_b,a_b) \right] \sqrt {n_b} \; ,\\
 \mu_{p0} \sqrt {n_p}
= \left[ - {\hbar^2 \over 2m_p} \nabla^2 +
 \ U_p +  \ \mu_p(n_p,a_f) \right] \sqrt {n_p} \; ,\label{eq11}
\eeqa
showing the agreement  between hydrodynamical  and nonlinear equations 
for 
superfluid bosons and fermions. 
Here $\mu_{b0}$ and $ \mu_{p0}$ are the chemical potentials for bosons 
and fermions in the stationary state.

{Although Eq. (\ref{eq11}) is written in terms of Fermi pair density 
$n_p$ (and we use this equation in the following), an equivalent equation 
for Fermi density $n_f$ (=2$n_p$) can be written as 
\beqa
 \mu_{f0} \sqrt {n_f}
= \left[ - {\hbar^2 \over 8m_f} \nabla^2 +
 \ U_f +  \ \mu_f(n_f,a_f) \right] \sqrt {n_f} \; ,\label{eq12} \\
\mu_f(n_f,a_f) =\frac{\hbar^2}{2m_f}(3\pi^2 n_f)^{2/3}g
(n_f^{1/3}a_f)\; ,
\eeqa
where we have used $\mu_{p0}=2\mu_{f0}$, $U_p=2U_f,$ and 
$\mu_p(n_p,a_f)=2\mu_f(n_f,a_f).$ }

Equations  
(\ref{nlse-b}) and (\ref{nlse-p}) are the Euler-Lagrange
equations of the following Lagrangian density
\beqa \label{lag}
{\cal L}_j &=& {i\hbar\over 2} \left(
\Psi_j^* {\partial \over \partial t} \Psi_j
- \Psi_j {\partial \over \partial t} \Psi_j^* \right)
- {\hbar^2\over 2  m_j} | \nabla \Psi_j |^2 \nonumber\\
&-&  U_j({\bf r}) |\Psi_j|^2 -  {\cal E}_j(n_j,a_j)
|\Psi_j|^2 \; ,
\eeqa
with $j=b,p$, (here $a_p=a_f$, the Fermi-Fermi
 scattering
length)
where
\beq
{\cal E}_j(n_j,a_j) = {1\over n_j} \int_0^{n_j} \mu_j(n'_j,a_j) \ dn'_j 
\;,
\eeq
is the bulk energy per particle of the superfluid.

Finally,  we introduce a Lagrangian density for Bose-Fermi interaction 
to
(\ref{lag})
of the form
\beq\label{lag2}
{\cal L}_{bp}= G_{bp}
|\Psi_b|^2 |\Psi_p|^2,
\eeq
where $G_{bp}={4\pi\hbar^2a_{bf}}/{m_{bf}}$, and  $m_{bf}$ is the
Bose-Fermi reduced mass and $a_{bf}$ the
Bose-Fermi scattering length.

{We note that there are no 
coupling terms between the 
hydrodynamical equations (\ref{eq4}) and (\ref{eq4a}) involving the 
velocity of two superfluid components. The only coupling in the 
Lagrangian density (\ref{lag2}) involves the density of the two 
superfluid components. However, the present hydrodynamical  equations 
(\ref{eq4}) and (\ref{eq4a})   are 
a special 
case of the three-fluid (two superfluids and a normal fluid) 
hydrodynamical formulation of Landau and Khalatnikov \cite{landau3} 
(appropriate for 
the $^4$He-$^3$He mixture). 
The  Galilei-invariant Landau-Khalatnikov formulation in its general 
form   
also has a coupling term involving two superfluid velocities (not 
considered in this paper), as discussed \cite{landau4}
by
Andreev and Bashkin  and also by Ho and Shenoy. 
However, the present static model defined by Eqs. (\ref{eq10}) and 
(\ref{eq11}) is obtained by setting all velocities equal to zero, and 
hence, these terms depending on superfluid velocities would leave the 
present model unchanged.}

{In this paper we consider  $a_{bf}$ to have  not too 
large positive values only (moderately repulsive Bose-Fermi 
interaction.) 
There are other domains of 
Bose-Fermi interaction  which require special attention. 
If this interaction is attractive  and strong enough, 
then the
ground state is very different, e.g., the  composite fermions $\tilde
{f}=bf$ are created at higher temperatures and at  zero temperature  we 
have Cooper
pairs (or local pairs) of the type $<\tilde {f}\mbox{  }\tilde {f}>$
consisting of two elementary fermions and two elementary bosons  
\cite{kagan}.}

The complete Lagrangian density given by Eqs. (\ref{lag}) and
(\ref{lag2}) leads to the following set of coupled equations for the
Bose-Fermi-Fermi mixture made of 
superfluid bosons and two-component superfluid fermions 
\beqa  \label{Q1}
i \hbar {\partial \over \partial t} \Psi_b 
&=& \left[ - {\hbar^2 \nabla^2
\over 2m_b} + 
U_b + \mu_b(n_b,a_b)+G_{bp}|\Psi_p|^2 \right] \Psi_b \; , \nonumber
\\ 
\\
\label{Q2}
i \hbar {\partial \over \partial t} \Psi_p
&=& \left[ - {\hbar^2 \nabla^2\over 2m_p}  + 
U_p + \mu_p(n_p,a_f)+G_{bp}|\Psi_b|^2 \right] \Psi_p \; ,  \nonumber \\
\eeqa
valid from weak-coupling to unitarity for both bosons and fermions. 
The coupled set of equations (\ref{Q1}) and (\ref{Q2}) with bulk 
chemical potentials  $\mu_b(n_b,a_b)$ and $\mu_p(n_p,a_f)$ given by Eqs. 
(\ref{cpbose}) and (\ref{fx}) (with $\nu=1.1$) and 
Eqs. (\ref{cpfermi}) and (\ref{gx})
[with $\delta= 20\pi/(3\pi^2)^{2/3}, \kappa=\delta/(1-\zeta), 
\zeta=0.44$], respectively,  is the US equation 
for the interacting superfluid Bose-Fermi mixture and 
is the required set.  However, when the bulk 
chemical potentials $\mu_b(n_b,a_b)$ and $\mu_p(n_p,a_f)$ 
are approximated using Eqs. (\ref{MGP}) and (\ref{MDF}), the set of 
equations (\ref{Q1}) and (\ref{Q2}) will be termed modified 
Gross-Pitaevskii density-functional
(MGPDF) equation including some correction for medium values of 
scattering lengths $a_b$ and $a_f$. Finally, if we approximate 
$f(x)=4\pi x$ and $g(x)=1$ in Eqs. (\ref{Q1}) and (\ref{Q2}) with 
 Eqs. (\ref{fx}) and
(\ref{gx}), the resultant model is termed Gross-Pitaevskii 
density-functional (GPDF) model. We present a comparative numerical 
study of these three models $-$ US, MGPDF, and GPDF $-$ in the next 
section.

\section{Numerical Results}

\begin{figure}[tbp]
{\includegraphics[width=.8\linewidth,clip]{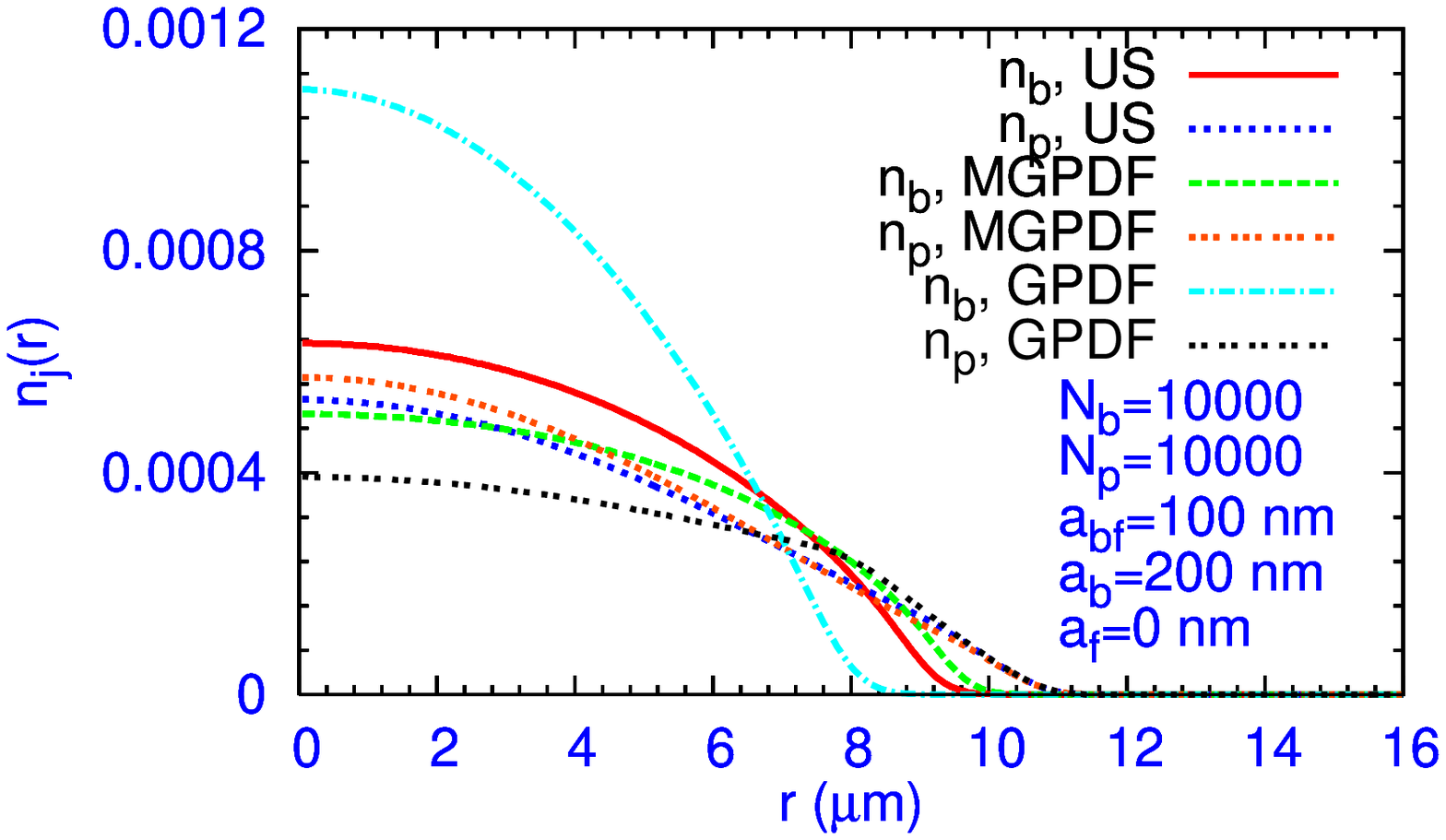}}
{\includegraphics[width=.8\linewidth,clip]{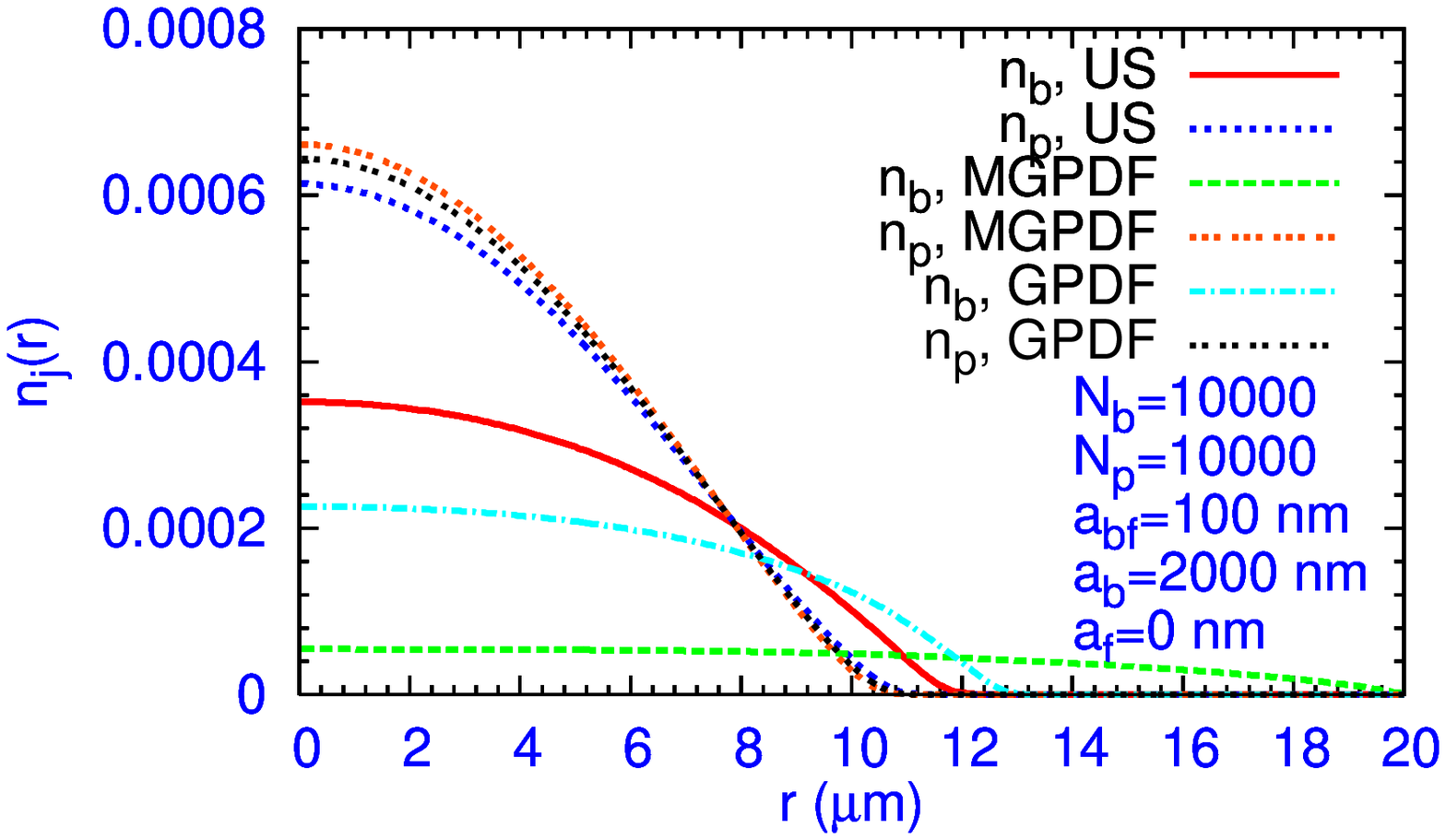}}
\caption{(Color online) The normalized densities $n_j(r)=|\Psi_j|^2$ 
with $j=b,p$ from a solution of Eqs.  (\ref{Q1})
and
(\ref{Q2}) for the US, MGPDF, and GPDF models with $N_b=10000, 
N_p=10000, a_{bf}=100$ nm, $a_f =0$, and (a) $a_{b}=200$ nm, and (b)  
$a_{b}=2000$ nm.  In this plot $\int n_j({\bf r}) d^3 {\bf r}=1.$}
\label{fig5}
\end{figure}

\begin{figure}[tbp]
{\includegraphics[width=.8\linewidth,clip]{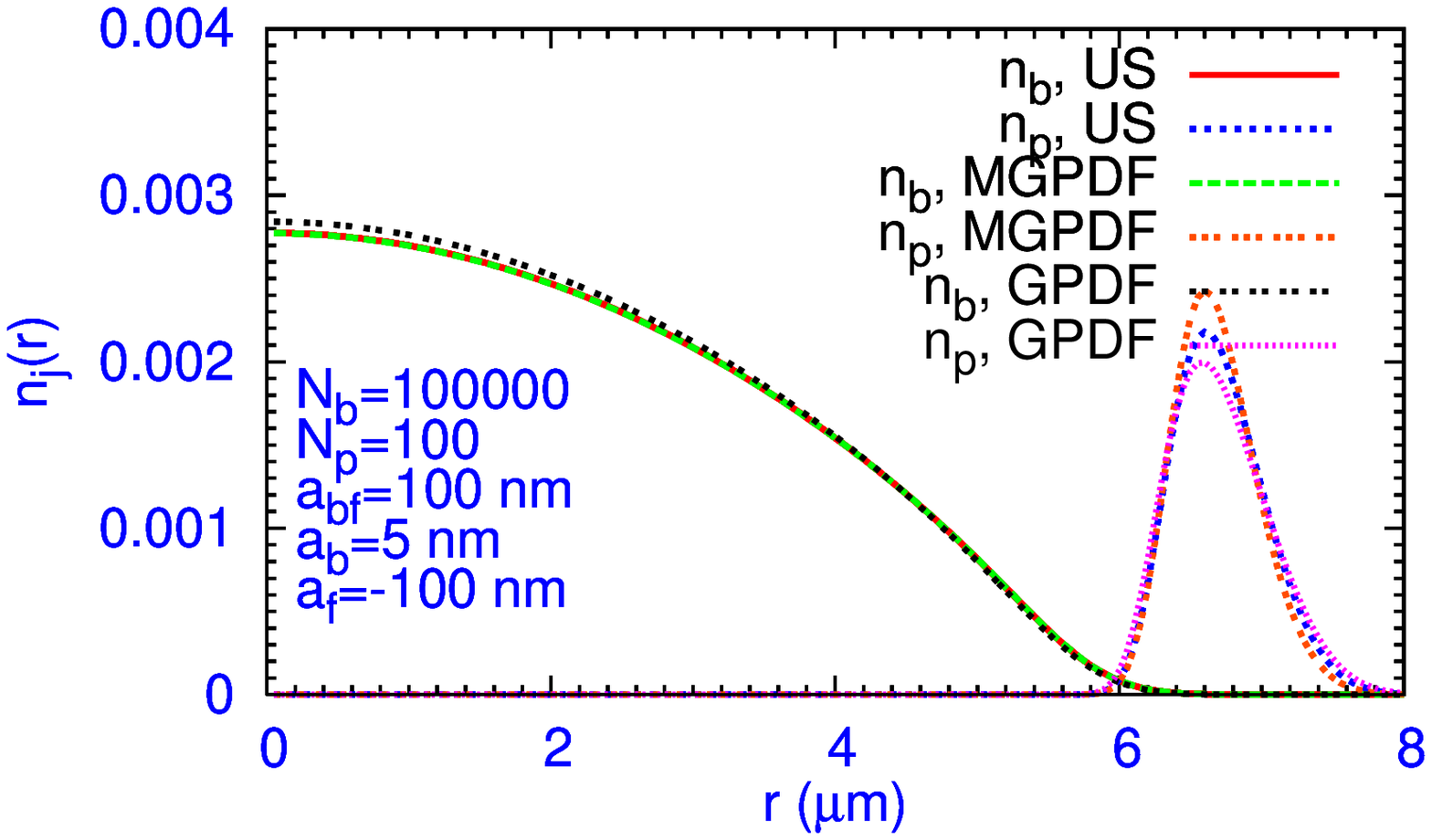}}
{\includegraphics[width=.8\linewidth,clip]{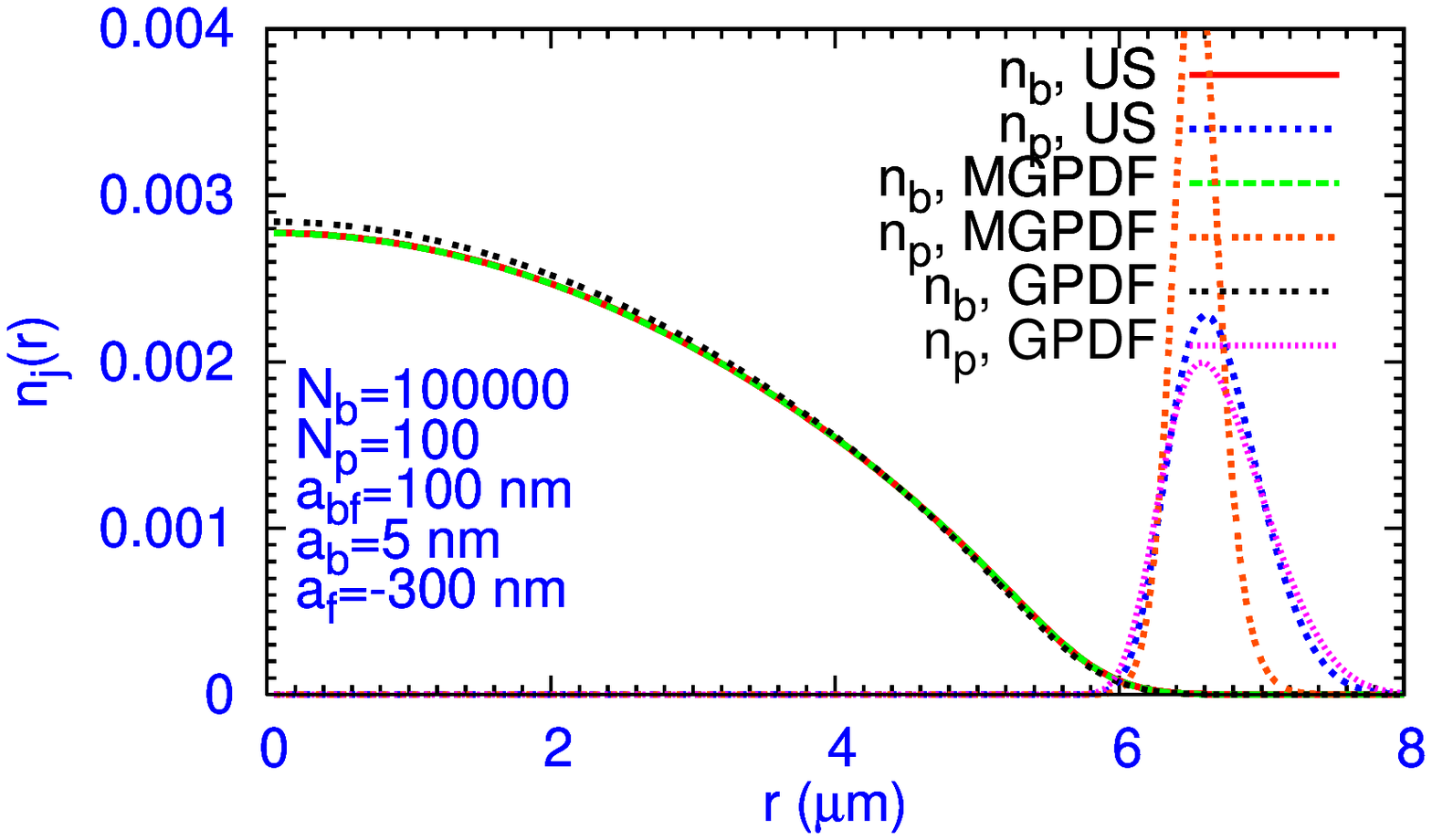}}
\caption{(Color online) The normalized densities $n_j(r)=|\Psi_j|^2$ 
with $j=b,p$ from a solution of Eqs.  (\ref{Q1})
and
(\ref{Q2}) for the US, MGPDF, and GPDF models with $N_b=100000, 
N_p=100, a_{bf}=100$ nm, $a_b =5$ nm, and (a) $a_{f}=-100$ nm, and (b)  
$a_{bf}=-300$ nm.  In this plot $\int n_j({\bf r}) d^3 {\bf r}=1.$}
\label{fig6}
\end{figure}

Here we solve numerically the coupled set of US Eqs.  (\ref{Q1}) 
and
(\ref{Q2}) with bulk
chemical potentials  $\mu_b$ and $\mu_p$ given by Eqs. (\ref{fx}) and
(\ref{gx}) and compare the results so obtained with those of the 
MGPDF and GPDF models.
We numerically solve the US, MGPDF, and GPDF  partial differential 
equations  by
discretizing them by the semi-implicit Crank-Nicholson algorithm with
imaginary time propagation \cite{num,sala-numerics,cn}.
The space and time steps used in
discretization were typically 0.05 and 0.001 respectively.

Although we are not interested here in fitting experimental data, we 
perform numerical calculation for variables for a possible  
experimental 
set up \cite{modugno}
of $^{87}$Rb-$^{40}$K mixture in  spherical harmonic traps satisfying
$U_b=m_b\omega_b^2 r^2/2= m_f\omega_f^2 r^2/2=U_f$   with 
an 
oscillator length   $l=\sqrt{\hbar/{(m_b\omega_b)}}= 1$  $\mu$m for 
$^{87}$Rb atoms. The oscillator length for $^{40}$K is modified 
according to the trap. 
The fermion pair trap  
in Eq.  (\ref{Q2}) is  
$U_p= m_f\omega_f^2 r^2$.

In Figs. \ref{fig5} (a) and (b) we plot densities for the superfluid 
Bose-Fermi 
mixture as calculated from Eqs. (\ref{Q1}) and (\ref{Q2}) for 
$N_b=10000, N_p=10000, a_f=0,  a_{bf}= 100$ nm with (a) $a_b=200$ nm 
and (b)   $a_b=2000$ nm for the US, MGPDF, and GPDF models. For small 
$a_b$ 
both MGPDF, and GPDF models are good approximations to the bosonic 
density of the 
US model 
(not shown in figure). But as $a_b$ increases, the MGPDF model is a 
better 
approximation to the bosonic density of the 
US model than the GPDF model as can be seen from Fig.  
\ref{fig5} (a). However, for very large $a_b$, neither the MGPDF nor the 
GPDF model provides the saturation of bulk chemical potential for 
bosons as in the 
US model. Consequently, the MGPDF, and GPDF models acquire large bosonic 
nonlinearity compared to the US model and hence produce bosonic density 
extending to larger region in space.  
As $a_b$ is increased, the  MGPDF model yields  
rapidly  divergent bosonic nonlinearity (compared to the GPDF model)  
and hence produces    bosonic density
 inferior to the GPDF  model 
as 
can be 
seen in Fig. \ref{fig5} (b).

\begin{figure}[tbp]
{\includegraphics[width=.8\linewidth,clip]{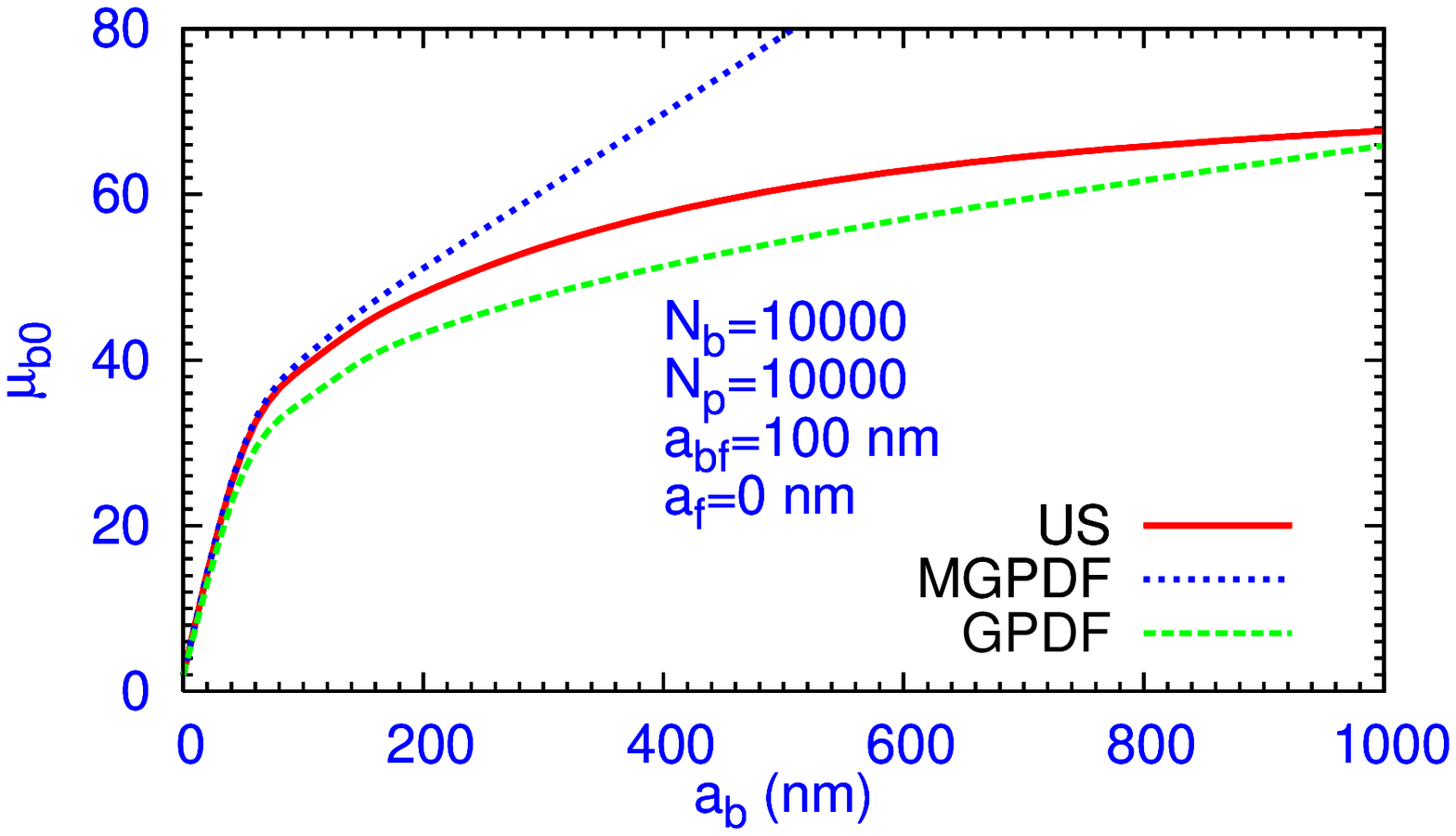}}
{\includegraphics[width=.8\linewidth,clip]{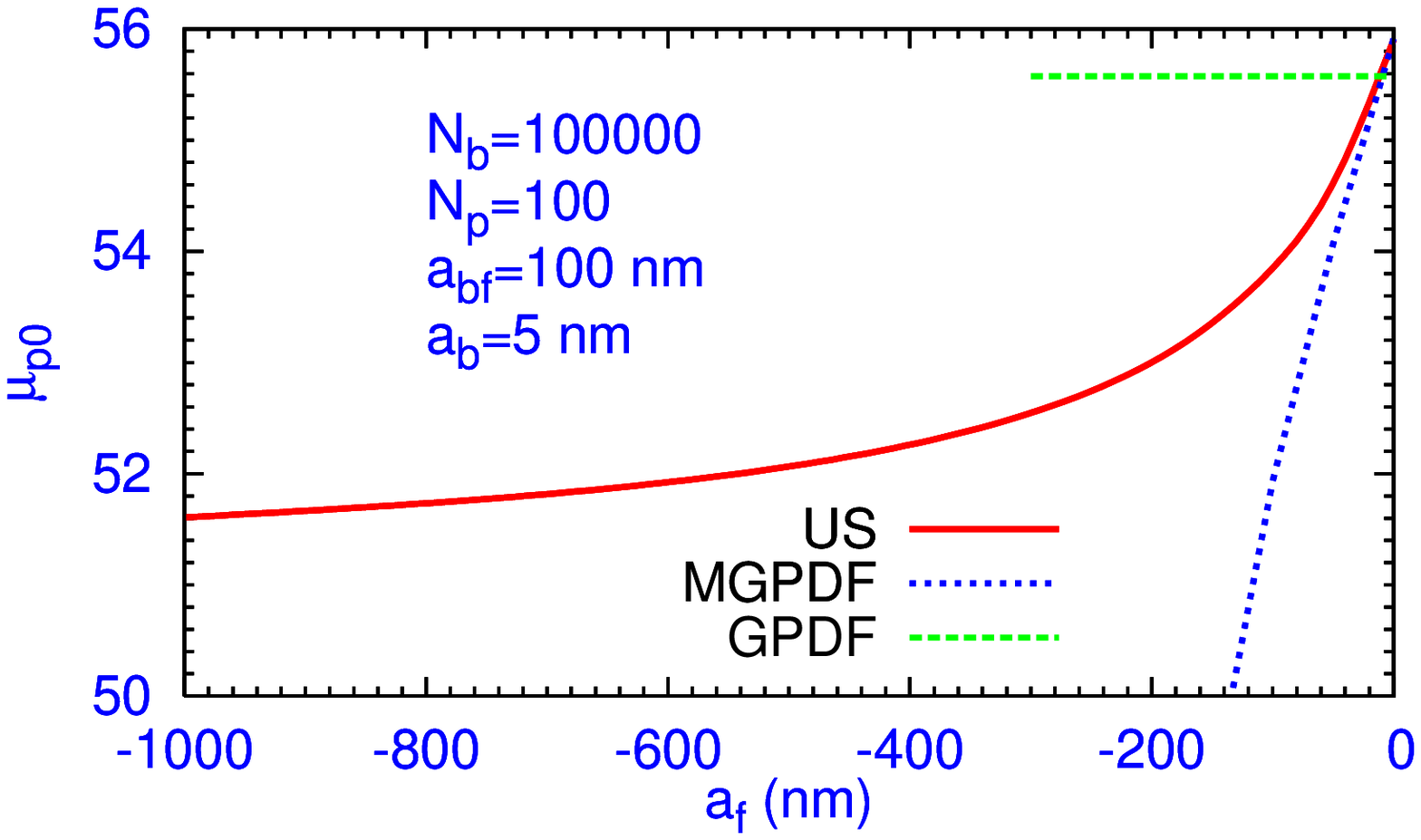}}
\caption{(Color online) (a) Chemical potential $\mu_{b0}$ vs. $a_b$ 
of Eqs. (\ref{eq10}) and (\ref{eq11})
for the US, MGPDF, and GPDF models
for  $N_b=N_p=10000, a_f=0,  a_{bf}=100$ nm and (b) chemical potential 
$\mu_{p0}$  vs. $a_f$
 for the US, MGPDF, and GPDF models
for 
$N_b=100000, N_p=100,  a_b=5$ nm,  $  a_{bf}=100$ nm, respectively. The 
chemical potentials are expressed in oscillator unit $\hbar\omega_b$.}
\label{fig7}
\end{figure}

In Figs. \ref{fig6} (a) and (b) we plot superfluid densities for 
$N_b=100000, N_p=100, a_{bf}=100$ nm, $a_b=5$ nm with (a) $a_f=-100$ nm 
and (b) $a_f=-300$ nm for the US, MGPDF, and GPDF models.
For small $|a_f|$ the fermionic densities of 
both MGPDF, and GPDF models are good approximation to the US model. 
But as $|a_f|$ increases, the GPDF model continues with BCS nonlinearity 
for $a_f=0$, whereas 
the fermionic nonlinearity of the MGPDF model acquires excessive 
attraction.  This is clear from Figs.  \ref{fig6} (a) and (b) showing a 
peaked fermionic density for the  MGPDF model 
for large $|a_f|$ corresponding to increased 
attraction. 
For very large $|a_f|$, the MGPDF 
model does not provide the saturation of fermionic bulk chemical 
potential as in the
US model.  We note that compared to the study of Figs. \ref{fig5} with 
comparable number of bosons and fermions, in the study of Figs. 
 \ref{fig6} we have much reduced number of fermions compared to bosons 
and the system has undergone a mixing-demixing transition 
\cite{md1,md} expelling the fermions from the central to the peripheral 
region. 

Next we calculate the chemical potentials $\mu_{b0}$ and $\mu_{p0}$
of the stationary superfluid 
Bose-Fermi
mixture as described by Eqs. (\ref{eq10})  and (\ref{eq11}). In Figs. 
\ref{fig7} (a) and (b) we plot these chemical potentials under the 
situations 
described in Figs. \ref{fig4}  and \ref{fig5}, e.g., in the 
(a) weak-coupling BCS limit when $a_b$ varies from a small to a large 
value and (b) weak-coupling GP limit when $|a_f|$  varies from a small 
to a large 
value, respectively. In the first case the fermionic chemical 
potential $\mu_{p0}$ remains practically constant over the full range 
of variation of $a_b$ 
and 
we plot only the bosonic chemical 
potential $\mu_{b0}$ in Fig. \ref{fig7} (a). In the second case 
bosonic chemical
potential $\mu_{b0}$ remains practically constant over the full range
of variation of $a_f$
and
we plot only the fermionic chemical
potential $\mu_{p0}$.   
From Fig. \ref{fig7} (a) we find that, although for small values of 
$a_b$, the MGPDF model produces results for $\mu_{b0}$ in close 
agreement 
with the US model, for medium values of $a_b$ the GPDF model produces 
results closer to the US model.   However, for larger values of $a_b$
the US model should show saturation, whereas the GPDF chemical potential 
$\mu_{b0}$ should increase monotonically with $a_b$ and should be larger 
than the the chemical potential of the US model (not shown in figure).   
From Fig. \ref{fig7} (b) we find that none of the MGPDF or GPDF models 
produces reasonable result for $\mu_{p0}$ except for very small values 
of $|a_f|$. The  GPDF model does not have a dependence on $a_f$ and 
hence 
produces a constant result for $\mu_{p0}$.

\section{Conclusion}

The dynamics  of cold trapped atoms (both bosons and fermions) should be 
correctly handled by a proper treatment of many-body dynamics or through 
a field-theoretic formulation. However, both approaches could be much 
too complicated to have advantage in phenomenological application. This 
is why mean-field approaches are widely used for phenomenological 
application. In the weak-coupling limit, for bosonic atoms the dynamics 
is handled through  the 
mean-field GP equation \cite{stringa1}, whereas for fermionic atoms the 
LDA (local density approximation) formulation  is 
widely used \cite{stringa2}. The LDA approach (often called the 
Thomas-Fermi 
approximation) lacks the kinetic energy gradient term (quantum 
pressure term). Here we introduce a quantum pressure term consistent 
with the correct phase-velocity relation of the superfluid fermions 
assuring the Galilei-invariance of the model \cite{sala-new}. 
This modified LDA equation for fermions and the GP equation for bosons, 
both valid in the weak-coupling limit, 
are both shown to be equivalent to the hydrodynamical  equation for 
superfluid flow.   
In the strong-coupling unitarity limit as the bosonic and fermionic 
scattering lengths $a_b$ and $|a_f|$  tend to infinity,  both the 
bosonic and fermionic interactions are to exhibit saturation due to 
constraints of quantum mechanics. We introduce proper unitarity 
saturation effects in these equations.  In addition in our Bose-Fermi 
formulation we introduce a not too strong contact interaction between 
bosons and fermions. The resultant superfluid Bose-Fermi dynamics is 
described by a set of coupled partial differential equation, 
which is the principal result of this paper. In 
the time-dependent form with a Bose-Fermi interaction 
this set is described by Eqs. (\ref{Q1}) and 
(\ref{Q2}) with the stationary version without Bose-Fermi 
interaction given by Eqs. (\ref{eq10}) and 
(\ref{eq11}).   

The present model for the superfluid Bose-Fermi mixture is termed the 
US model, valid in both weak- and strong-coupling unitarity limits for 
bosons and fermions. We also consider a model called the GPDF model 
where the boson dynamics is handled by the GP Lagrangian and the fermion 
dynamics is handled by the LDA formulation with a proper kinetic energy 
gradient term. Finally, we also consider another model called the MGPDF 
model including lowest order correction to the GPDF model due to the 
non-zero bosonic and fermionic scattering lengths $a_b$ and $|a_f|$.
However, the  MGPDF
model does not provide a saturation of the nonlinear interaction in the 
unitarity limit when $a_b$ and $|a_f|$ tend to infinity. In our 
numerical studies of a $^{87}$Rb-$^{40}$K mixture in a spherical 
harmonic trap, we find that for medium interactions the MGPDF
model produces improved  results. However, for stronger interactions the 
full US model should be used. Although, slightly complicated for 
analytic applications, the present US model is no more complicated than 
the usual GP and LDA equations for numerical treatment and is expected 
to find wide applications in future studies of Bose-Fermi superfluid 
mixture. 

\acknowledgments

We thank D. Blume for kindly providing additional data of FNMC 
calculation \cite{FNMC2}.
SKA has been partially supported by the FAPESP and CNPq of Brazil.  
LS has been partially supported by the Fondazione CARIPARO 
and GNFM-INdAM.

\end{document}